\documentclass[twocolumn,secnumarabic,amssymb,nobibnotes,aps,prl,nofootinbib]{revtex4-1}
\usepackage[english]{babel}
\usepackage{graphicx}
\usepackage{amsmath}
\usepackage{amsfonts}
\usepackage[usenames,dvipsnames]{color}
\usepackage{hyperref}
\usepackage{wasysym}  
\usepackage{rotating}
\usepackage{fancyhdr}
\usepackage{cancel}
\usepackage{setspace} 
\usepackage{tikz}
\usepackage{colortbl}
\usepackage{transparent}

\begin{document}

\author{Mathijs Janssen, Ben Werkhoven, and Ren\'{e} van Roij}
\affiliation{Institute for Theoretical Physics, Center for Extreme Matter and Emergent Phenomena,  Utrecht University, Leuvenlaan 4, 3584 CE Utrecht, The Netherlands}
\date{\today}

\title{Harvesting vibrational energy with liquid-bridged electrodes: thermodynamics in mechanically and electrically driven RC-circuits}

\begin{abstract}
We theoretically study a vibrating pair of parallel electrodes bridged by a (deformed) liquid droplet, which is a recently developed microfluidic device to harvest vibrational energy. The device can operate with various liquids, including liquid metals, electrolytes, as well as ionic liquids. We numerically solve the Young-Laplace equation for all droplet shapes during a vibration period, from which the time-dependent capacitance follows that serves as input  for an equivalent circuit model. We first investigate two existing energy harvesters (with a constant and a vanishing bias potential), for which we explain an open issue related to their optimal electrode separations, which is as small as possible or as large as possible in the two cases, respectively. Then we propose a new engine with a {\em time-dependent} bias voltage, with which the harvested work and the power can be increased by orders of magnitude at low vibration frequencies and by factors 2-5 at high frequencies, where frequencies are to be compared to the inverse RC-time of the circuit.
\end{abstract}

\maketitle

Small-amplitude oscillations are ubiquitous. Not only devices like fans, laundry machines and speakers vibrate, pretty much everything around us does. Converting these mechanical vibrations into electric energy could provide a valuable alternative to batteries in portable electronic devices which require only modest amounts of electric power \cite{roundy2005effectiveness}.  Moreover, powering remote sensors with vibrations could relieve the requirement of connection to the electricity grid.
Unfortunately, engines based on induction \cite{beeby2006energy} or piezo electricity \cite{anton2007review} are not well suited for these small-scale applications since their power performance rarely surpasses the 0.1W range \cite{krupenkin2011reverse}. In search of a promising alternative, variable-capacitance engines have received considerable interest in recent years  \cite{meninger2001vibration, boisseau2012electrostatic}. 

Variable-capacitance engines operate by cyclically (dis)charging electrodes at alternating high (low) capacitance. Net electric energy is harvested during a cycle because the charging stroke occurs at a lower potential than the discharging stroke.
The change in capacitance can be caused by a mechanical stimulus as in the case of vibrational-energy harvesters, but also by a change in the properties of the dielectric or electrolyte material. 
Examples of the latter include electrolyte-filled nanoporous supercapacitors where variable capacitance is achieved by changing electrolyte concentration (in capacitive mixing) \cite{brogioli2009extracting, hamelers2013harvesting}, or temperature (in capacitive thermal energy extraction) \cite{hartel2015heat}, or combinations thereof \cite{Janssen:2014aa, ahualli2014temperature}. 

\begin{figure}
\centering
\begin{picture}(8.44,5.8)
\put(0.1,0.0){\includegraphics[width=8.2cm]{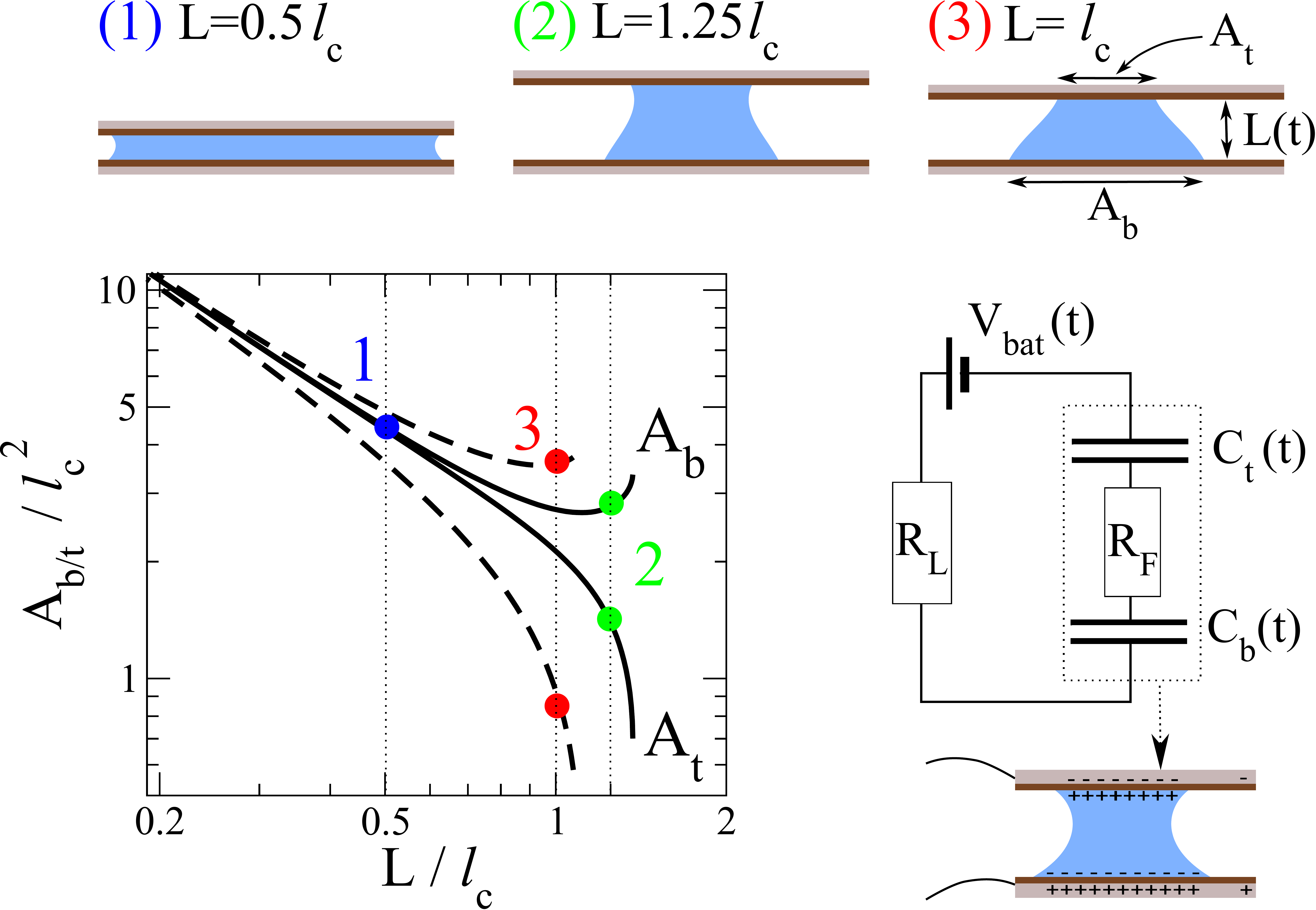}}
\put(0.0,5.5){{\bf (a)}}
\put(0.0,3.9){{\bf (b)}}
\put(5.3,3.9){{\bf (c)}}
\end{picture}
\caption{\scriptsize{(a) Liquid bridges between electrodes separated over $L=0.5l_{c}, 1.25l_{c}, l_{c}$ (1,2,3) as found using the Young-Laplace eqn~(\ref{eq:YoungLaplace}). Here $l_{c}$ is the capillary length, typically in the mm regime. The droplets have a volume $\mathcal{V} =\frac{2\pi}{3} l_{c}^{3}$ and equal contact angles $\theta_{b/t}=70^\circ$ (1,2) and unequal contact angles $\theta_{b}=70^\circ, \theta_{t}=105^\circ$ (3).
(b) The plate separation $L$ dependence of the bottom and top contact areas $A_{b}, A_{t}$, for a droplet of the same volume, and $\theta_{b/t}=70^\circ$ (lines), and $\theta_{b}=70^\circ, \theta_{t}=105^\circ$ (dashed). 
(c) Equivalent circuit model of a liquid-bridge vibrational energy harvester. }}
\label{fig1}
\end{figure}
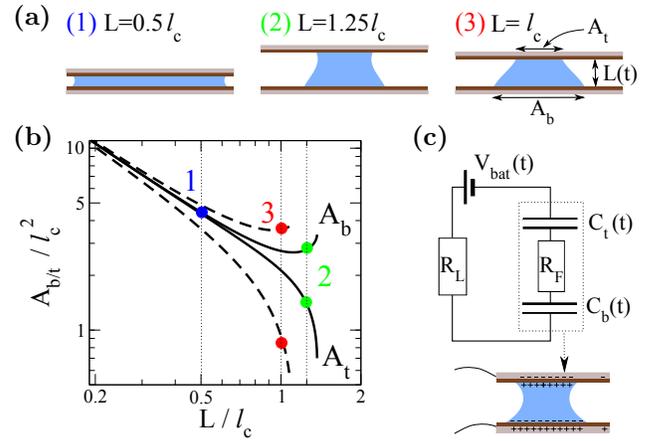

Variable-capacitance engines driven by mechanical energy typically consist of air-filled parallel-plate capacitors connected to a battery, where the capacitance is modified either by varying the plate separation or the lateral plate overlap \cite{miyazaki2003electric}. 
A key new development in these engines was recently realized by Krupenkin and Taylor  \cite{krupenkin2011reverse} who suggested to inject an array of small liquid droplets (Mercury and Galinstan) between the electrodes.  Charge on the capacitor's plates is now balanced by the polarization of a dielectric film whose thickness is orders of magnitudes smaller than the plate separation (see Fig.~\ref{fig1}). This leads to significantly larger capacitances and therefore to higher power densities of the order of $10^3$Wm$^{-2}$.  Due to the difference in permittivity between air and most liquids, most charges will accumulate at the droplet-electrode interface. When the liquid bridges are cyclically compressed and stretched by a varying electrode separation, the change in droplet shape leads to a different contact area and hence capacitance, which at a constant bias potential directly drives a current. 

To increase the suitability for practical applications, a simpler circuitry was subsequently proposed without the need for an external power source by Moon {\it et al} \cite{moon2013electrical}. Instead of an imposed bias potential, in this set-up the symmetry between the two planar electrodes is broken by a difference in contact angles at the liquid-air-electrode three-phase contact lines. Because electrode-specific surface reactions will always lead to finite (but small) charge densities on the electrodes, a current will flow when the plates are oscillating. However, the small charge densities lead to a low power production of the order of $\sim 10^{-3}$Wm$^{-2}$. An additional problem facing this set-up is that the water droplets used in this study tend to vaporize after a few oscillations, requiring a continuous injection. This problem was overcome by Kong {\it et al} \cite{kong2014ionic}  who investigated five imidazolium ionic liquids which, due to their low volatility and toxicity, do not require an air tight space. Since the properties of the ionic liquids can be adjusted by choosing different ions, they are ``designable materials" that allow one to tailor-make the droplets to fit any desired application. For example, operation outside the liquid temperature range of water is now accessible.

To the best of our knowledge, no systematic optimization study of the different engines w.r.t. to the mechanical driving has been performed so far. 
While both Moon {\it et al} \cite{moon2013electrical} and Kong {\it et al} \cite{kong2014ionic} showed that work output increases with increasing oscillation amplitude,  the average plate separation was fixed in both studies. 
Note that with increasing amplitude of the plate oscillations, at some point the assumptions underlying the theoretical analysis of the mentioned works will break down.
More specifically, in Ref.~\cite{krupenkin2011reverse} it was assumed that a sinusoidal driving of the plate separation leads to a sinusoidally varying droplet-surface contact area $A$ and hence capacitance. Conversely, the droplet engines with asymmetric contact angles \cite{moon2013electrical,kong2014ionic} have a sinusoidally varying capacitance {\it only} on the hydrophobic surface, since the contact line on the hydrophilic surface was assumed to be perfectly pinned. The pinning can for instance be due to surface heterogeneities but, as hypothesized by the authors, can also be attributable to the difference in contact angles. Regardless of its physical nature, pinning of the contact line will have to break down when the droplet is strongly compressed by large-amplitude oscillations (especially on low-hysteresis surfaces). 

To access the regime of large-amplitude oscillations, of special interest when optimizing the delivered power, in this Communication we go beyond the previous analysis by explicitly solving the Young-Laplace equation that describes the droplet shape, assuming that droplets perfectly conform to Young's law at the contact line. 
The droplet profiles obtained serve as input for an equivalent circuit model, whose solutions can conveniently be displayed in the charge-potential representation. Since these are conjugate thermodynamic variables, the enclosed area during an operation cycle represents the  work delivered in one oscillation period. 
We will show that the engine with bias potential thrives in the small plate-separation regime, due to a large relative variable capacitance. Conversely, the asymmetric plate engine will be found to thrive at large plate separation, because the hydrophilic plate contact area, which contributes negatively to the performed work, is ``effectively" pinned. The mentioned pinning is not related to surface heterogeneities, but is actually emerging in our idealized model.  
Decreasing the plate separation breaks the pinning, and hence breaks the asymmetry between the plates.

Having confirmed and explained the results of Ref's  \cite{krupenkin2011reverse,moon2013electrical, kong2014ionic}, we then propose a new engine with a {\it time-varying bias potential}. At the cost of introducing an increased complexity in the circuitry, we show that this engine harvests substantially more electric energy than the droplet engines considered so far. While the enhancement is quantitative in the high-frequency domain, in the regime of low oscillation frequency we find a qualitative difference, attributed to the now finite work per cycle performed in this regime.

Consider two parallel electrode plates separated by a distance $L$, bridged by a droplet of volume $\mathcal{V}$ as visualized in Fig.~\ref{fig1}(a). The bottom plate is located in the plane $z=0$ and the top plate in $z=L$, with $z$ anti-aligned with the direction of gravity.  
The shape of the cylindrically symmetric droplet as described by the unit normal $\hat{n}$ to the fluid-air interface is determined by the Young-Laplace equation \cite{batchelor2000introduction}, together with the boundary conditions set by Young's law at the droplet-plate interface, 
\begin{eqnarray}
\gamma\vec{\nabla}\cdot \hat{n}&=&\Delta p-g \Delta \rho z, \label{eq:YoungLaplace}\\
\gamma_{lg}\cos{\theta_{b/t}}&=&\gamma_{sl}-\gamma_{sg}\label{eq:Young}.
\end{eqnarray}
in terms of a pressure drop $\Delta p$ that acts as a Lagrange multiplier for $\mathcal{V}$, gravitational acceleration $g$, droplet mass density $\rho$, and surface tensions $\gamma$ between gas ($g$), liquid ($l$) and solid ($s$). By choosing different dielectric coatings one can access different tensions $\gamma_{sl}$ and $\gamma_{sg}$ on either electrode plates, leading to different contact angles $\theta_{b/t}$ at the bottom and top plate. Note that throughout this Communication the subscripts (b/t) refer to quantities associated to the bottom and the top plate, respectively.

Hydrodynamic effects, which at typical parameters start to become important when the electrodes oscillate at frequencies above $f\sim 50$Hz, are ignored \cite{moon2013electrical,kong2014ionic}. 

Typical droplet profiles resultant of eqn~(\ref{eq:YoungLaplace})-(\ref{eq:Young}) are shown in Fig.~\ref{fig1}(a). Varying the plate separation at fixed $\mathcal{V}$, for each droplet profile one measures the contact areas $A_{b/t}$ of the droplet with the bottom and top plate. In this way we obtain the relation between plate separation and the respective contact areas. This relation is shown in Fig.~\ref{fig1}(b)
 in units of the capillary length $l_{c}=\sqrt{\frac{\gamma_{lg}}{g\Delta\rho}}$, which is the typical length scale of the problem ($l_{c}=2.7$mm for water and air at room temperature).
For small plate separation $L\ll l_{c}$ (configuration 1 in Fig.~\ref{fig1}(a)) the droplet is essentially a pancake-shaped cylinder such that $A_{b}\simeq A_{t}\simeq \frac{\mathcal{V}}{L}$.
When the plate separation is increased to the order of the capillary length $L\sim l_{c}$ (details depend on relative contact angles), the droplet starts to cave in under the influence of gravity, such that $A_{b}$ levels off while $A_{t}$ shrinks much faster than $\sim L^{-1}$ (configuration 2 and 3 in Fig.~\ref{fig1}(a)) .
In Fig.~\ref{fig1}(b) where we plot $A_{b}$ and $A_{t}$ as a function of $L$, we observe that small variations of the plate separation can affect $A_{t}$ by an order of magnitude, while $A_{b}$ remains roughly constant. This effective pinning of the hydrophilic bottom plate contact area, as mentioned in \cite{moon2013electrical, kong2014ionic}, is {\it emerging} rather than {\it imposed}. It is attributable both to the difference in contact angles {\it and} the effect of gravity. Gravity alone is in principle enough to observe this effect, in the case of equal contact angles we observe in Fig.~\ref{fig1}(b) a similar effect, but at larger plate separation. 

A liquid-bridge variable-capacitance engine is constructed by connecting this droplet to a battery at potential $V_{\rm bat}(t)$ and a load of resistance $R_{\rm L}$ over which the harvested energy will be dissipated (see Fig.~\ref{fig1}(c)). The electrodes, both coated with a dielectric layer of microscopic thickness $d$ and dielectric constant $\epsilon_{d}$, will acquire {\it net} charges $q_{b/t}(t)$. These charges are a combination of {\it induced} charges $\mp q(t)$, together with {\it fixed} charges $Q_{b/t}$, such that we write  $q_{b}(t)=Q_{b}-q(t)$ and $q_{t}(t)=Q_{t}+q(t)$. The fixed charges can be due to surface-specific reactions \cite{moon2013electrical} but are typically small, such that any non-zero bias-potential effectively leads to a dominance of the induced charges $|q(t)|\gg Q_{b/t}$.

The capacitance $C_{b/t}$ on each electrode is an in-series connection of a dielectric capacitor with an electric double-layer (EDL) capacitor. Within the Helmholtz picture of the EDL the resultant harmonic mean reads
\begin{equation}\label{eq:capacitances}
C_{b/t}=\epsilon_{0} A_{b/t}[L] \left(\frac{d}{\epsilon_{d}}+\frac{\kappa^{-1}}{\epsilon_{l}}\right)^{-1},
\end{equation}
where the Debye screening length $\kappa^{-1}$ sets the typical width of the EDL of ions dissolved in the droplet of dielectric constant $\epsilon_{l}$. In the absence of a dielectric layer one can resort to more accurate descriptions of the EDL which are well established \cite{kornyshev2007double, hartel2015fundamental}. 
The capacitor is driven mechanically; we focus on a sinusoidal oscillation in the plate separation $L(t)=L_{0}+\Delta L \sin \omega t$, with average plate separation $L_{0}$, oscillation amplitude $\Delta L$, and oscillation frequency $\omega$. Consequently, the surface areas $A_{b/t}[L]$, found through the Young-Laplace eqn~(\ref{eq:YoungLaplace}), are time varying, leading on the basis of eqn~(\ref{eq:capacitances}) to a time-varying capacitance.
We find the potential of each of the electrodes in terms of the net charges and capacitances with $V_{b/t}\equiv\frac{q_{b/t}}{C_{b/t}}$. 
This definition in terms of the charges present on the electrode plates requires an additional minus sign when splitting the capacitor potential $V_{\rm cap}=-V_{b}+V_{t}$.
Kirchoff's voltage law now gives the following governing equation
\begin{eqnarray}
R\frac{dq(t)}{dt}-\frac{Q_{b}-q(t)}{C_{b}(t)}+\frac{Q_{t}+q(t)}{C_{t}(t)}&=&V_{\rm bat}(t),\label{eq:kirchoffgeneral}\\
\frac{q(0)}{C_{tot}(0)}&=&V_{\rm bat}(0),\label{eq:initialcondition}
\end{eqnarray}
where with $R=(R_{\rm L}+R_{\rm F})$ we included an internal fluid resistance $R_{\rm F}$ in the capacitor due to viscous energy losses associated with ion currents within the fluid. Here we also defined the total capacitance as $C_{tot}(t)^{-1}\equiv C_{b}(t)^{-1}+C_{t}(t)^{-1}$.  

The work delivered to the load $W_{\rm L}$ and average power $\langle P\rangle$ performed during one oscillation can be found by integrating the instantaneous power $P(t)=R_{\rm L}\left(\frac{dq(t)}{dt}\right)^{2}$ over one oscillation period $T=2\pi \omega^{-1}$,
\begin{eqnarray}\label{eq:work}
W_{\rm L}=\langle P\rangle T&=&\int^{T}_{0} dt P(t)=
\int^{T}_{0} dt R_{\rm L}\left(\frac{dq(t)}{dt}\right)^{2}.
\end{eqnarray}
Interestingly, this is easily connected to thermodynamics, where delivered work is related to areas enclosed by cycles in planes of conjugate thermodynamical variables. Rewriting the integral in eqn~(\ref{eq:work}) as $\int^{T}_{0} dt(..)=\oint dq \frac{dt}{dq}(..)$, we can split off the energy dissipated over the internal droplet resistance $R_{\rm F}$ and  insert Kirchoff's law in the remaining terms,
\begin{eqnarray}\label{eq:thermodynamicderivation}
W_{\rm L}&=&\oint dq V_{\rm bat}-\oint dq V_{\rm cap}-\oint dq R_{\rm F}\frac{dq}{dt}\label{eq:thermodynamicderivation2}, \\ 
&\equiv&W_{\rm bat}+W_{\rm har}+W_{\rm lost}.
\end{eqnarray}
The term $W_{\rm lost}$ represents viscous and/or Ohmic losses associated with the ion current within the droplet, which in general will be small  because for most systems $R_{\rm F}\ll R_{\rm L}$.
From the First Law $d\mathcal{F}=-FdL+V dq$  (with $\mathcal{F}(q,L,\mathsf{T})$ the free energy of the system at fixed temperature $\mathsf{T}$, and with $F$ the force between the plates) follows the definition of
electrostatic work $dW\equiv-V dq$ done by the capacitor. 
This provides the interpretation that the work delivered to the load as found in eqn~(\ref{eq:thermodynamicderivation}) is a combination of work performed by the battery $W_{\rm bat}$, together with energy harvested from the conversion of mechanical to electric energy $W_{\rm har}$.  It is the latter term, and {\it not} the total dissipated energy over the load $W_{\rm L}$, that should be optimized for the purpose of energy harvesting. 

In the remainder of this Communication we study the steady-state solutions of eqn~(\ref{eq:kirchoffgeneral}) in three specific cases: {\it case 1} a constant bias potential $V_{\rm bat}(t)=V_{0}$ with vanishing fixed charges $Q_{b/t}=0$ \cite{krupenkin2011reverse};  {\it case 2} no external battery power, with fixed charges $Q_{b/t}\neq0$; and  {\it case 3} a time-varying battery potential $V_{\rm bat}(t)$ with vanishing fixed charges $Q_{b/t}=0$.
When calculating quantities with physical dimensions, we use the following realistic values: $R_{\rm L}=10^{6}\Omega$; a (Teflon) dielectric coating of $d=400$nm,  $\epsilon_{d}=2.1$; a (ionic) liquid droplet $\epsilon_{l}=11$, $R_{\rm F}=40 \Omega$, $\gamma_{lg}=72$mN m$^{-1}$, $\rho=0.997$g cm$^{-3}$ such that $l_{c}=2.7$mm, $\kappa^{-1}=0.05$nm   \cite{moon2013electrical, kong2014ionic}.
For these values, the capacitance eqn~(\ref{eq:capacitances}) is dominated by the capacitance of the dielectric layer, because $\frac{d}{\epsilon_{d}}\gg\frac{\kappa^{-1}}{\epsilon_{l}}$. This inequality also holds for a water-like droplet with a Debye length in the nm regime (much lower salt concentration) and larger dielectric constant.  
Moreover, for these parameters, Lippman corrections $\gamma_{eff}=\gamma_{lg}-\frac{CV_{\rm cap}^{2}}{2A}$ to the contact angles can be ignored \cite{klarman2011model,buehrle2003interface} provided that the bias voltage is smaller than $V_{\rm cap}\ll\sqrt{2A\gamma_{lg}C^{-1}}\approx 58$V.  

\begin{figure}
\centering
\begin{picture}(8.44,12.0)
\put(0.2,6.1){\includegraphics[width=8.0cm]{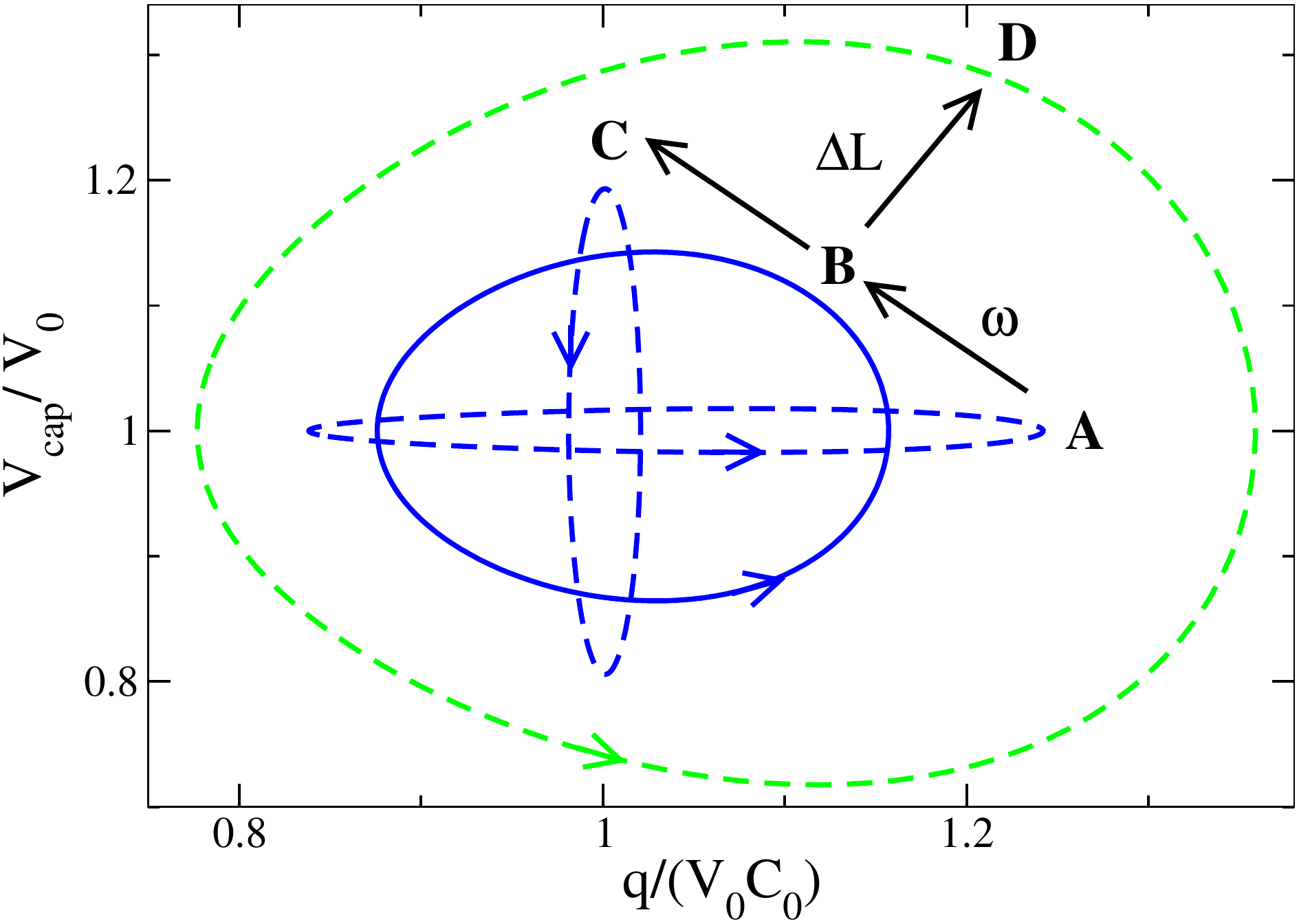}}
\put(0.2,0.0){\includegraphics[width=8.0cm]{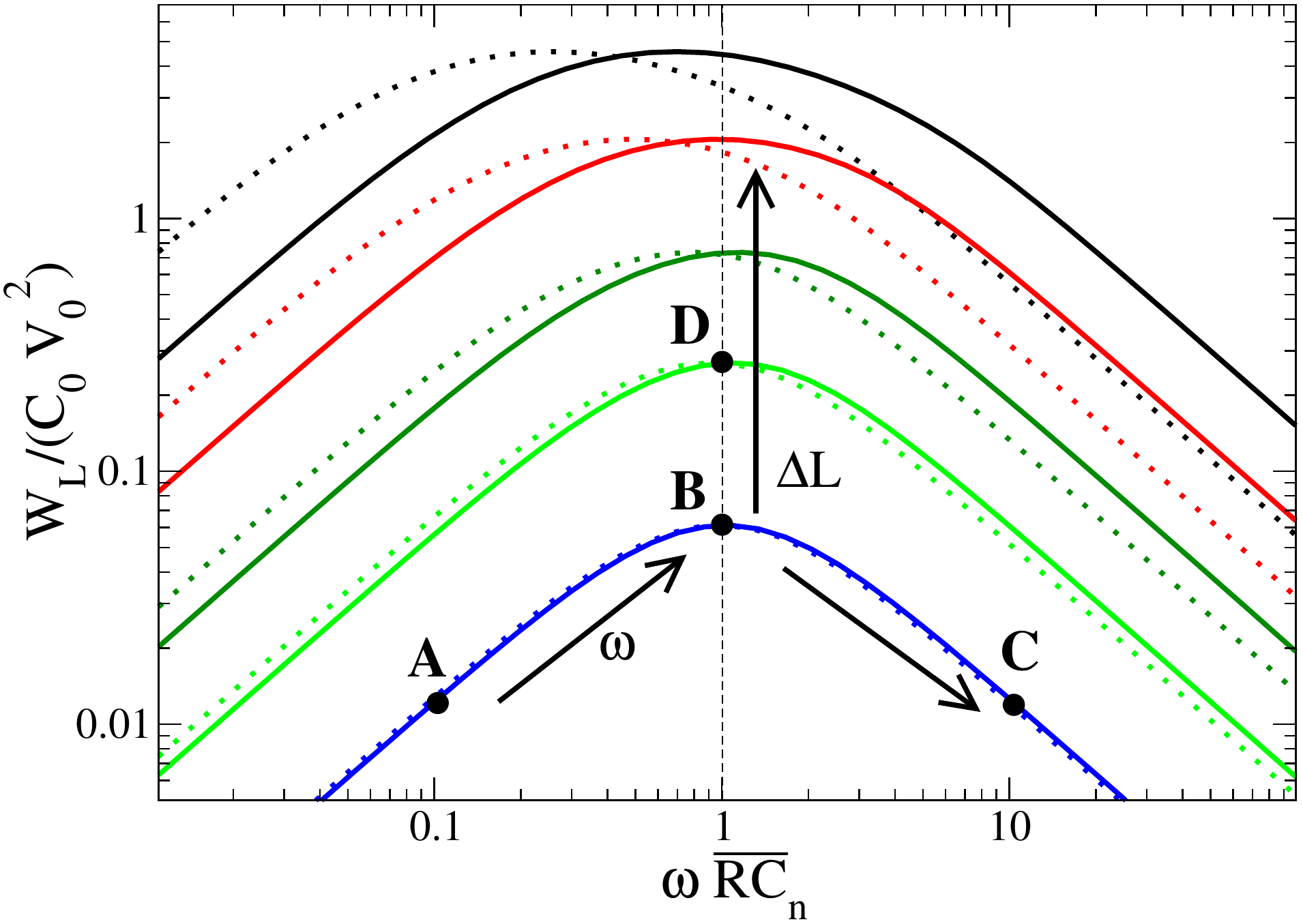}}
\put(0.0,11.66){{\bf (a)}}
\put(0.0,5.56){{\bf (b)}}
\end{picture}
\caption{\scriptsize (a) Parametric plot of a full cycle in the charge-capacitor potential representation, at average plate separation for $L_{0}/l_{c}=0.25$ for three different driving frequencies $\omega \overline{RC}_{1}=\{0.1,1,10\}$ (A,B,C) and two oscillation amplitudes $\Delta L/L_{0}=0.2, 0.4$ (Blue lines (A,B,C), Green line D)
\scriptsize (b) Performed work $W_{\rm L}$ at  $L_{0}/l_{c}=0.25$ for $\Delta L/L_{0}=0.2, 0.4, 0.6, 0.8, 0.95$ (from bottom to top). The frequency domain was rescaled with $\overline{RC}_{1}$ ({\it dotted}) and $\overline{RC}_{4}$ ({\it lines}). The four cycles {\bf A,B,C,D} shown in part (a) are indicated in part (b).}\label{fig:parametricplot}
\end{figure}

Using our Young-Laplace results, typical solutions to eqn~(\ref{eq:kirchoffgeneral}) in {\it case 1} are displayed in the plane of conjugate charge-capacitor potential variables ($q,V_{\rm cap}$) in Fig.~\ref{fig:parametricplot}(a). 
For $L_{0}/l_{c}=0.25$ and $\Delta L/L_{0}=0.2$ we show cycles for three angular frequencies $\omega \overline{RC}_{1}=\{0.1,1,10\}$ denoted with {\bf A,B,C}, respectively. Here the angular frequencies $\omega$ of the mechanical driving of $L(t)$ are made dimensionless with the average RC-time as defined by a generalized mean
\begin{equation}\label{eq:averageRC}
\overline{RC}_{n}=\displaystyle{\left[\frac{1}{T}\int_{0}^{T} dt (RC_{tot})^{n}\right]^{1/n}}.
\end{equation}
This definition with $n=1$ was introduced in Ref.~\cite{moon2013electrical} and sets the typical response time of the electronic system. Note that in this case $W_{\rm bat}=V_{0}\oint dq=0$ and therefore drops out of eqn~(\ref{eq:thermodynamicderivation}) such that $W_{\rm L}=W_{\rm har}$ in this case (this equality was also checked numerically). 
At the low frequency of $\omega \overline{RC}_{1}=0.1$ (cycle {\bf A}) the electronic system is nearly relaxed throughout the mechanical driving process, such that a small potential over the load only slightly affects the potential over the capacitor with respect to the bias potential. Conversely, at the high frequency $\omega \overline{RC}_{1}=10$ (cycle {\bf C}) when the electronic system cannot relax in response to the mechanical driving, there is a large spread in capacitor potential, with high currents but little charge is flowing between the capacitor plates.  
In between these eccentric frequencies lies an optimal ``resonance" frequency $\omega \overline{RC}_{1}\sim 1$ (cycle {\bf B}) for which the enclosed area in the ($q,V_{\rm cap}$) representation is maximal. We furthermore show a cycle {\bf D} at the resonance frequency but at a larger plate amplitude $\Delta L/L_{0}=0.4$. This cycle clearly encloses more area and hence harvests more energy.  
These conclusions are supported by Fig.~\ref{fig:parametricplot}(b) where we show the $\omega$-dependence of the work $W_{\rm L}$ for a variety of amplitudes $\frac{\Delta L}{L_{0}}$, where $\omega$ is rescaled with $\overline{RC}_{1}$ (dotted) and $\overline{RC}_{4}$ (full lines).
Interestingly, this figure shows a remarkable resemblance to the Bode magnitude plot of a traditional RC-circuit, especially for the generalized mean in eqn~(\ref{eq:averageRC}) with $n=4$ (see also Fig.~\ref{fig:traditionalRC} in the {\it supporting information}). The generalized mean with $n=1$ performs well for $\frac{\Delta L}{L_{0}}\ll1$, but deviates appreciably when this ratio becomes of order unity, where the variance $\overline{RC}^{2}_{2}-\overline{RC}^{2}_{1}$ also increases significantly. 
From Fig.~\ref{fig:parametricplot}(b) we furthermore see that this engine delivers more work for increasing $\frac{\Delta L}{L_{0}}$. In the limit $\frac{\Delta L}{L_{0}}\to1$, the contact area diverges when $L(t)\to0$ during a part of the cycle, leading to high values (and variation) of the capacitance. 

In a set-up with {\it no external power source}, $V_{\rm bat}\equiv0$ ({\it case 2}), electrode-specific reactions of the electrolyte absorbing to the plates can yet lead to net (equilibrium) charges $Q_{b}$ and $Q_{t}$ on the plates \cite{moon2013electrical}.
The initial condition eqn~(\ref{eq:initialcondition}) reduces to the relation between the equilibrium charges $\frac{Q_{b}}{C_{b}(0)}=\frac{Q_{t}}{C_{t}(0)}$, which ensures that the Kirchoff law is satisfied in mechanical equilibrium (no driving). 
We can thus rewrite eqn~(\ref{eq:kirchoffgeneral}) to
\begin{eqnarray}\label{eq:Kirchoffasym}
R\frac{dq(t)}{dt}+\frac{q(t)}{C_{tot}(t)}&=&Q_{b}\left(\frac{1}{C_{b}(t)}-\frac{C_{t}(0)}{C_{b}(0)C_{t}(t)}\right).
\end{eqnarray}
In the absence of an imposed bias potential, the r.h.s. of this equation, linear in the fixed charge $Q_{b}$,  takes the role of an effective ``time-varying voltage source". 
\begin{figure}
\centering
\begin{picture}(8.44,9.16)
\put(0.3,5.2){\includegraphics[width=3.18cm]{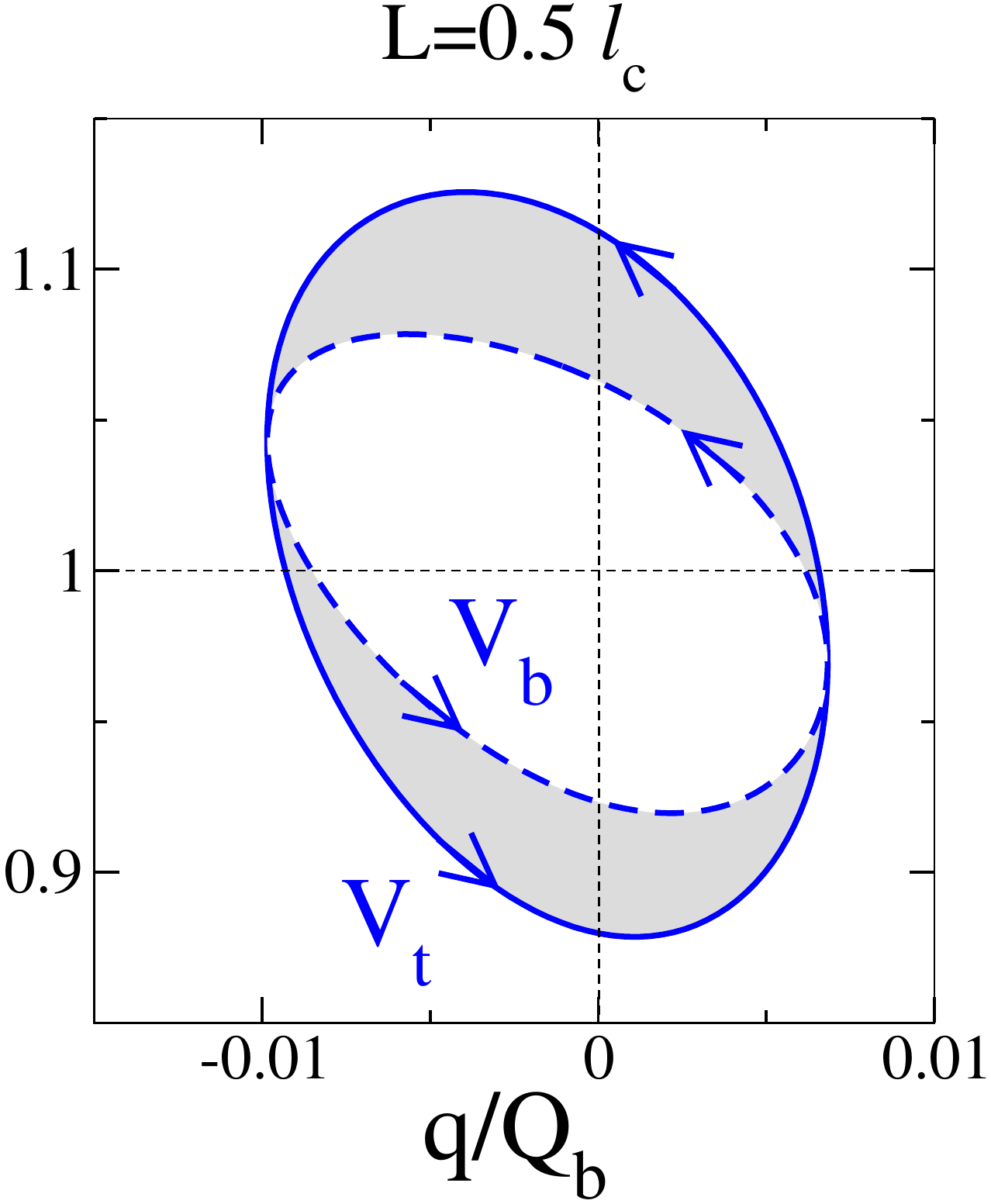}}
\put(3.5,5.2){\includegraphics[width=4.82cm]{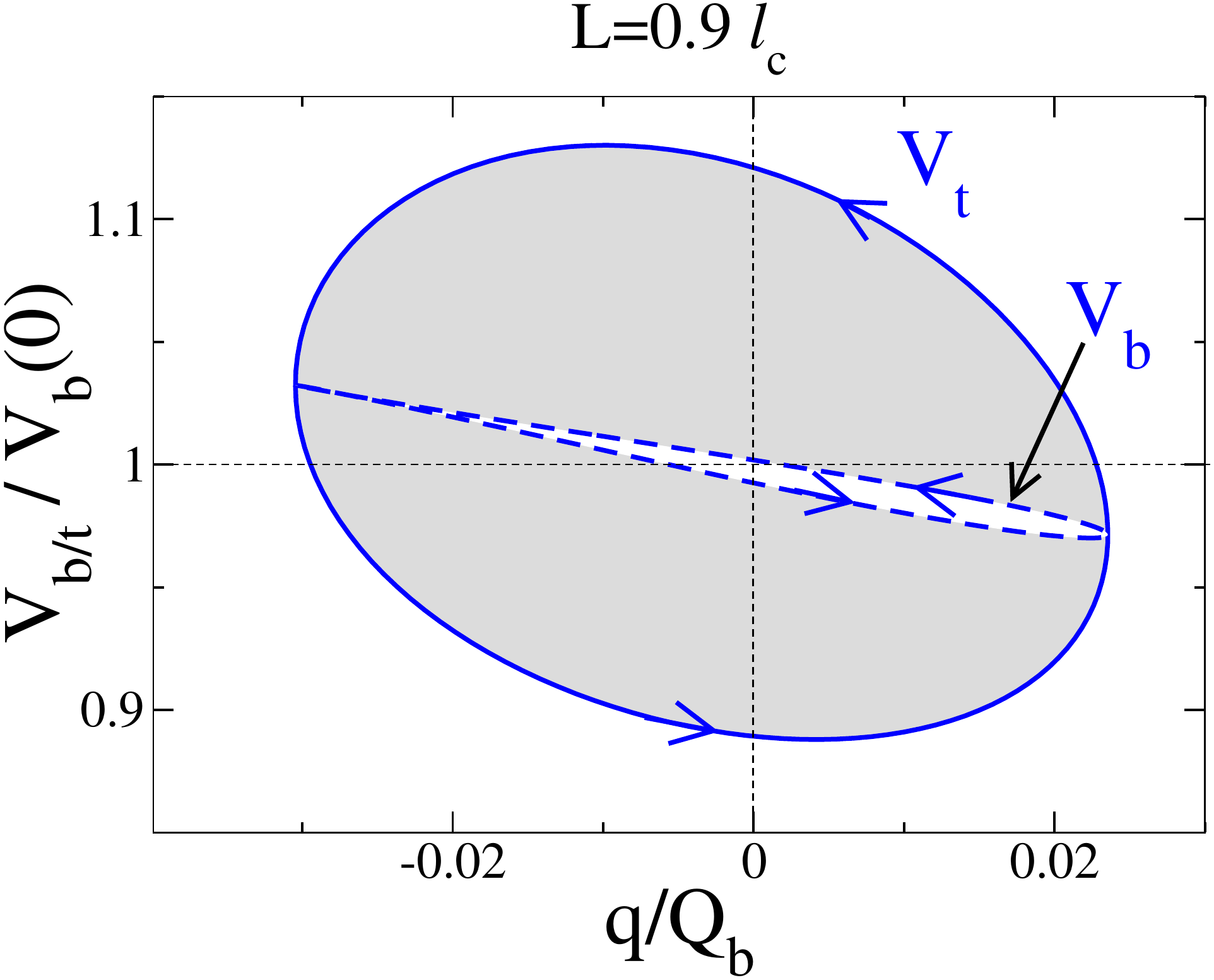}}
\put(0.36,0.0){\includegraphics[width=8cm]{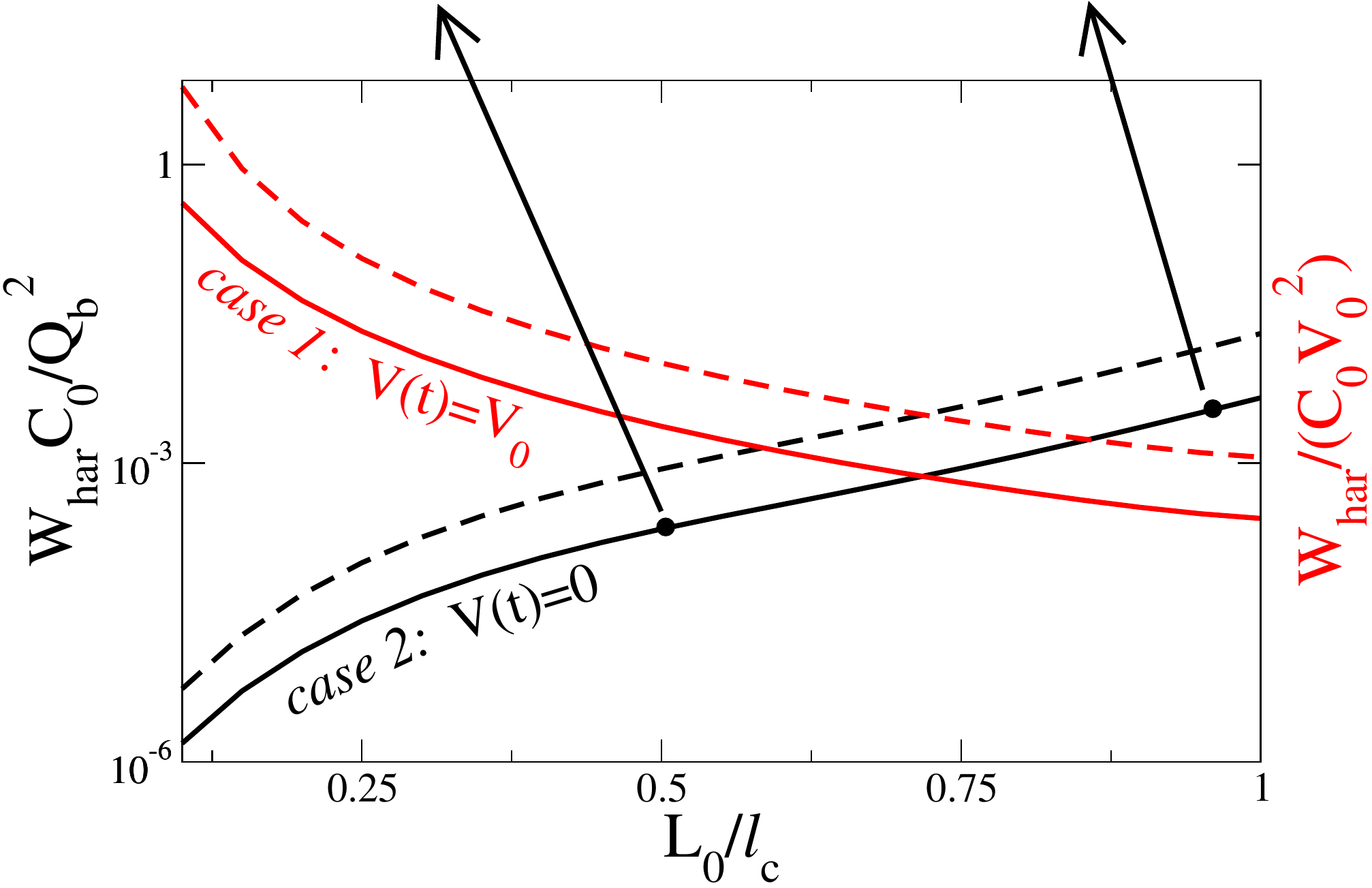}}
\put(0.0,9.06){{\bf (a)}}
\put(3.9,9.06){{\bf (b)}}
\put(0.0,4.56){{\bf (c)}}
\end{picture}
\caption{\scriptsize (a,b) Parametric plots in the charge-potential plane of the charging state of the oscillating liquid bridges for  $\Delta L/l_{c}=0.05$, $f=20$Hz, $R_{L}=30$M$\Omega$, fixed charges density $\sigma\equiv \frac{Q_{b}(0)}{A_{b}(0)}=4.18\cdot 10^{-5}$ Cm$^{-2}$, with $V_{b}$ (dashed) and $V_{t}$ (lines). 
Increasing the plate separation from $L_{0}=0.5l_{c}$ (a) to $L_{0}=0.9l_{c}$ (b) induces an effective pinning of the bottom plate and narrows the area enclosed by the inner line.
(c) $L_{0}$ dependence of the performed work per cycle at oscillation frequency $f=20$Hz and amplitudes $\Delta L/l_{c}=0.05, 0.1$ (lines, dashed), for asymmetric plates without bias voltage for $R_{L}=30$M$\Omega$ (black) and symmetric plates with bias voltage  $R_{L}=1$M$\Omega$ (red).}
\label{fig3:Ldependence}
\end{figure}
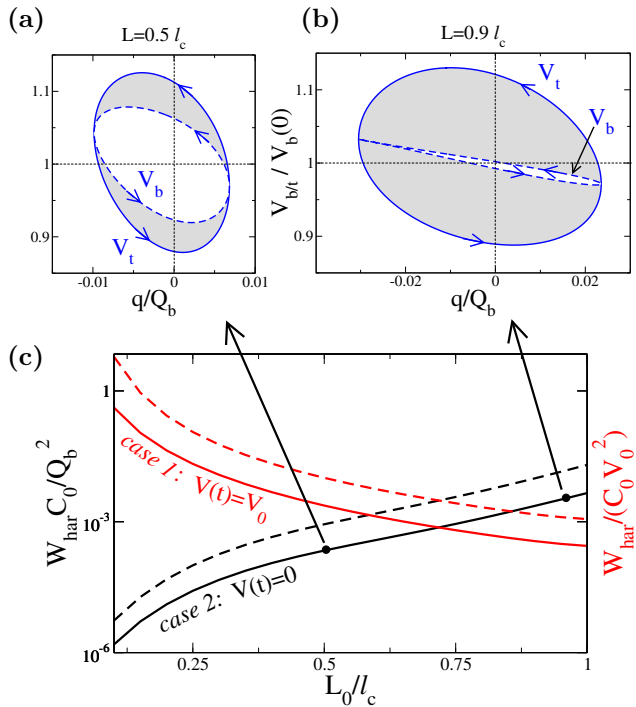
To investigate the relative contributions of both plates to the energy harvesting, in Fig.~\ref{fig3:Ldependence} we show the cyclic charging state of each electrode separately: in (a) for a relatively small electrode separation of $L_{0}=0.5l_{c}$, and in (b) for $L_{0}=0.9l_{c}$. In the absence of an external voltage source, the variables are now made dimensionless with the initial charge on the bottom capacitor $Q_{b}$ (see {\it supporting information} for details). 
In eqn~(\ref{eq:thermodynamicderivation}) we separate the contributions of each of the electrodes to find (ignoring viscous losses over $R_{\rm F}$),
\begin{eqnarray}\label{eq:workasym}
W_{\rm har}&=&\oint dq V_{b}-\oint dq V_{t}. 
\end{eqnarray}
Fig.~\ref{fig3:Ldependence}(a) and (b) clearly show that the net charge is dominated by initial charges $Q_{b/t} \gg |q(t)| \Rightarrow q_{b/t}(t)\approx Q_{b/t}$ for this case, 
such that the potentials $V_{b/t}$ appearing in eqn~(\ref{eq:workasym}) have the same sign. An extra minus sign due to the anti-clockwise nature of the charging cycles in the ($q,V_{\rm cap}$) representation leads to the conclusion that  the bottom plate does negative work. It then follows from eqn~(\ref{eq:workasym}) that the grey shaded area in Fig.~\ref{fig3:Ldependence}(a) and (b) represents the harvested work $W_{\rm har}$. This is to be contrasted with {\it case 1} for which $q_{b/t}(t)=\mp q(t)$ such that both   (comparable) contributions of the electrodes to the harvested work add up. 

As was mentioned in Ref.~\cite{moon2013electrical}, decreasing the average plate separation has an adverse effect on the harvested work. 
Within our model this can be interpreted on the basis of the asymmetry between bottom and top plate. This asymmetry, induced both by gravity and the different contact angles, is stronger for larger electrode separations (Fig.~\ref{fig1}(b)).
At large plate separations $L\sim l_{c}$
small variations of $L$  only affect the surface area of the hydrophobic top plate. The bottom plate, subject to an ``effective pinning", has a (roughly) constant contact area hence capacitance, and therefore the first term on the r.h.s. of eqn~(\ref{eq:workasym}) vanishes.
Conversely, decreasing $L_{0}$ will lift the asymmetry between the bottom and the top plate: the contact areas will now be similarly affected by variation of $L$. The advent of variable capacitance on the bottom plate is observed in Fig.~\ref{fig3:Ldependence}(a) as the ``opening up" of the white area enclosed by the dashed line, when compared to Fig.~\ref{fig3:Ldependence}(b) where the bottom electrode does not contribute. 
Due to the relative minus sign in eqn~(\ref{eq:workasym}), the now finite contribution of the bottom plate has a detrimental effect on the harvested power, which is seen in Fig.~\ref{fig3:Ldependence}(a) as a decrease in the grey shaded area. 

In Fig.~\ref{fig3:Ldependence}(c) we summarize (on a log-scale that spans many decades) the two opposing trends of decreasing/increasing work-output in {\it case1/case2} upon increasing the average spacing $L_{0}$.

The introduction of a {\it time-dependent bias potential} $V_{\rm bat}(t)$ ({\it case 3}) in principle could improve the performance of the droplet engine, for example
when the capacitor potential $V_{\rm cap}$ is lowered when the electrodes are charging and increased when electrodes discharge. 
Note that in this case eqn~(\ref{eq:thermodynamicderivation}) picks up a finite contribution from the battery ($\oint dq V_{\rm bat}\neq0$). 
We consider a square-wave bias potential $V_{\rm bat}(t)=V_{\rm sq}(t)\equiv V_{0}+\lim_{\xi\to\infty}\Delta V \tanh\left({\xi \sin{\left(\omega t+\delta\right)}}\right)$ (we use $\xi=30$), at the same angular frequency $\omega$ as the mechanical driving. The variable voltage source $V_{\rm sq}(t)$ is not only characterized by an average potential $V_{0}$ and an amplitude $\Delta V$, but also by a phase difference $\delta$ between the imposed mechanical vibration $L(t)$ and the applied potential $V_{\rm bat}(t)$, with respect to which we will optimize. 
\begin{figure}
\centering
\begin{picture}(8.44,16.2)
\put(0.25,10.3){\includegraphics[width=8.05cm]{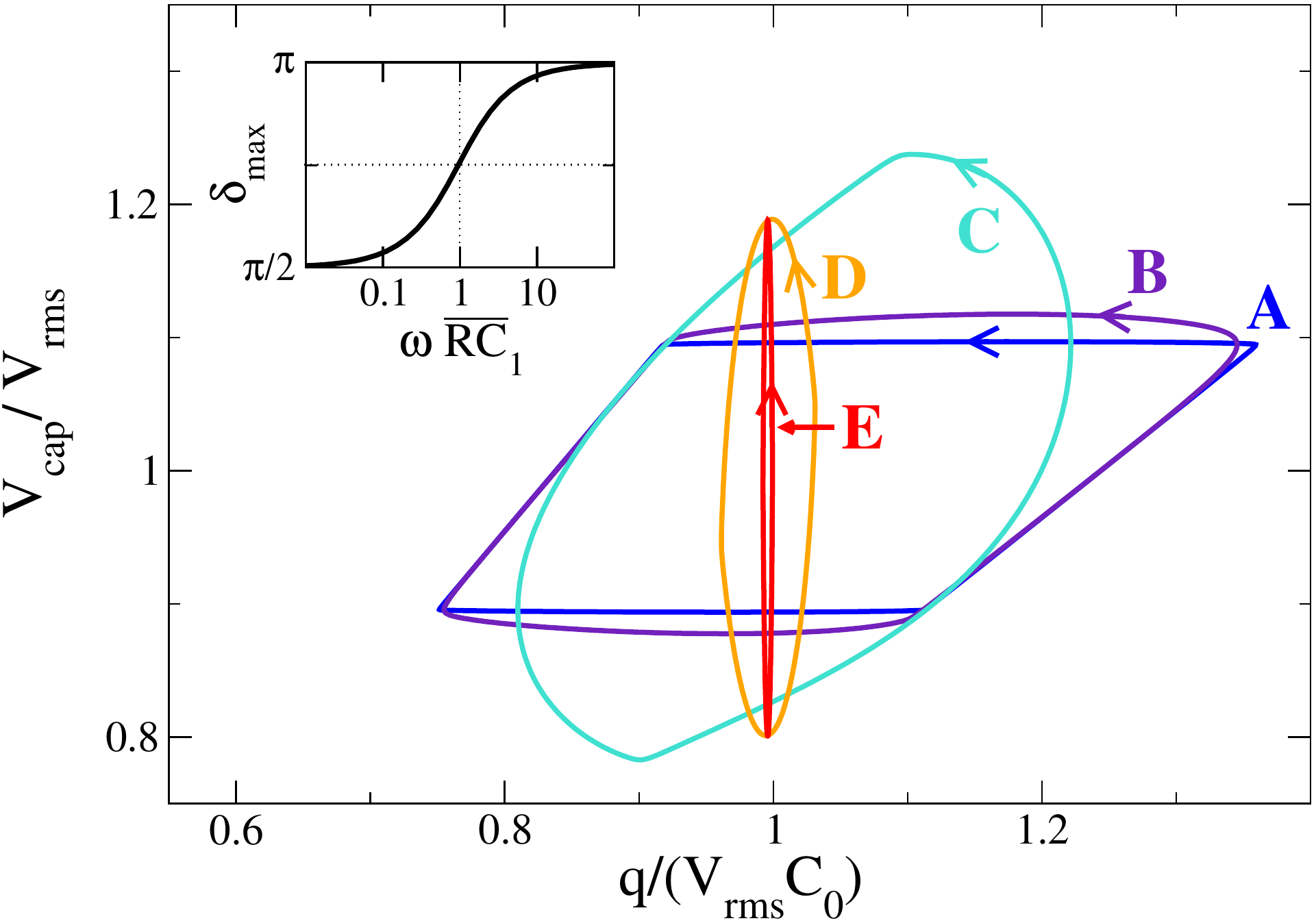}}
\put(0.2,4.8){\includegraphics[width=8.0cm]{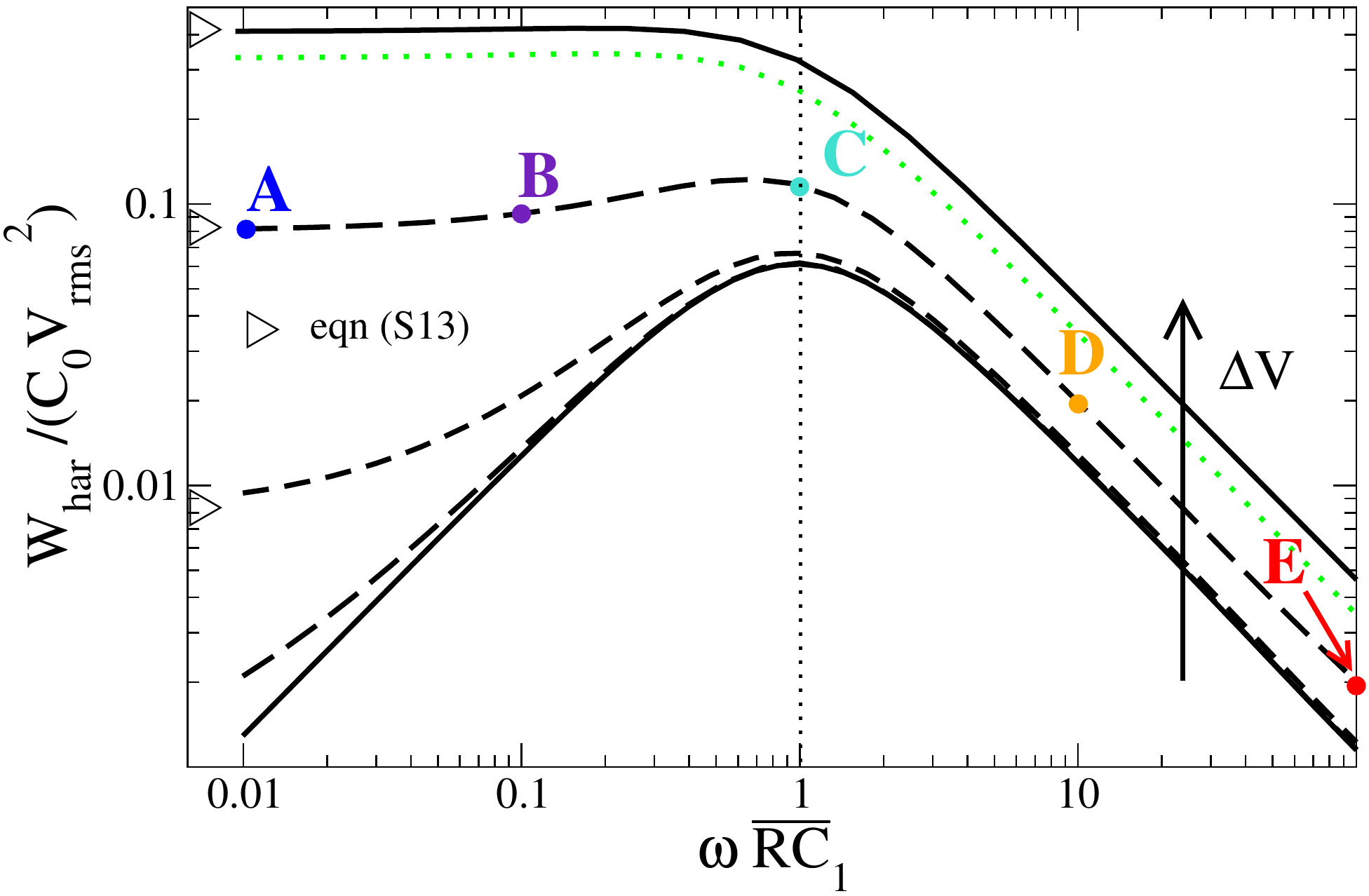}}
\put(0.2,0.0){\includegraphics[width=8.0cm]{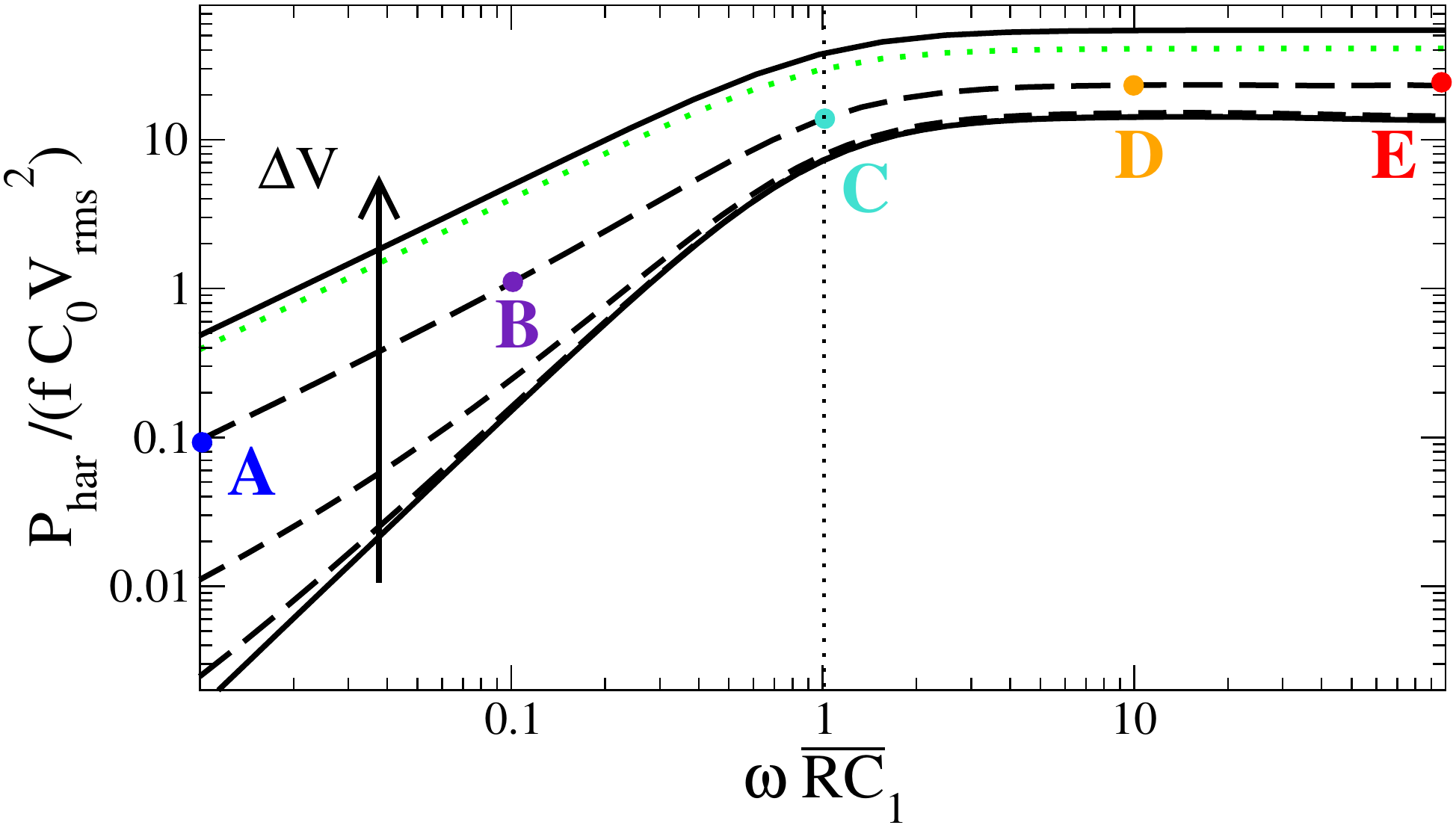}}
\put(0.0,15.66){{\bf (a)}}
\put(0.0,9.96){{\bf (b)}}
\put(0.0,4.56){{\bf (c)}}
\end{picture}
\caption{\scriptsize{(a) Charging cycles evaluated at their respective optimal phase differences $\delta_{\rm max}$ (inset) in the charge-potential representation for the square wave bias potential, for $\Delta V/V_{0}=0.1$ at $L_{0}/l_{c}=0.25$ and $\Delta L/l_{c}=0.05$. We show $\omega \overline{RC}_{1}=\{0.01, 0.1, 1, 10, 100\}$ ({\bf A-E}). 
(b) The cycle-averaged dimensionless harvested work as a function of $\omega \overline{RC}_{1}$, for several $\Delta V/V_{0}=\{0, 0.001, 0.01, 0.1, 1\}$ (black, from bottom to top) and $\Delta V/V_{0}=2$ (green, dotted). Note that $\Delta V/V_{0}=0$ is {\it case 1}. Cycles {\bf A-E} are highlighted as dots on the pink line.
(c) The cycle-averaged dimensionless power as a function of $\omega \overline{RC}_{1}$ for the same parameters. }}
\label{fig4:cycles}
\end{figure}
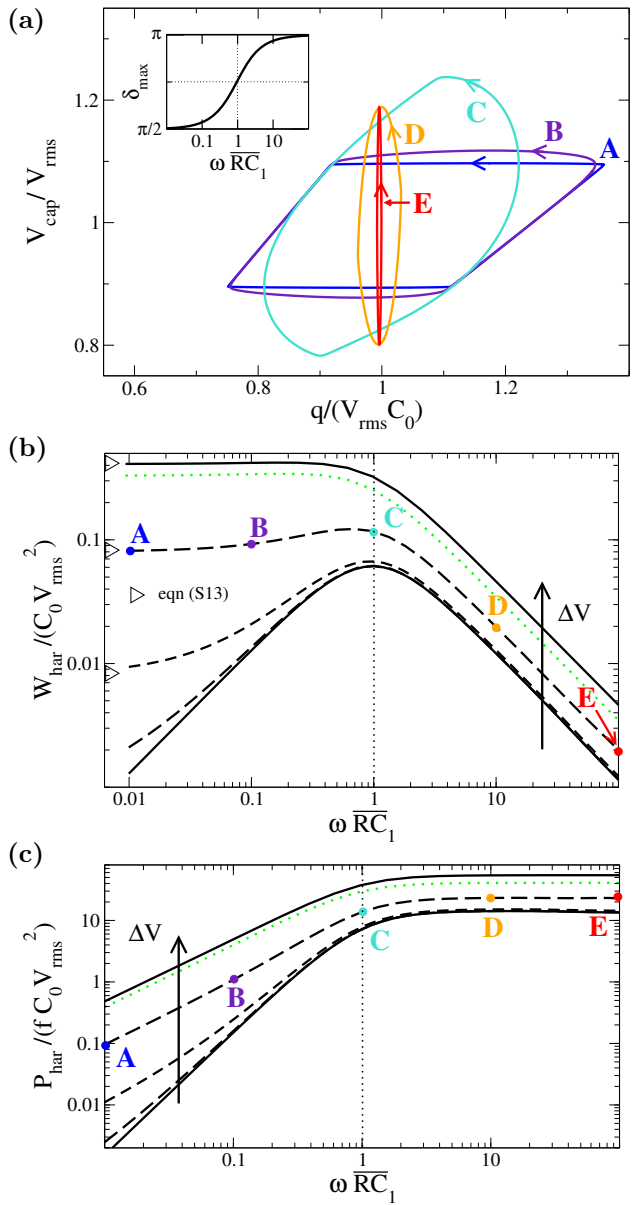

In Fig.~\ref{fig4:cycles}(a) we show the steady state of the charging/discharging cycle of the capacitor in the charge-voltage representation $(q,V_{\rm cap})$ for a set of frequencies $\omega \overline{RC}_{1}=\{0.01, 0.1, 1, 10, 100\}$ (denoted with {\bf A-E}), each evaluated at their respective $\delta_{\rm max}$, the value of $\delta$ for which $W_{\rm har}$ is maximized.  The voltages are rescaled with the root-mean-square bias potential $V^{2}_{\rm rms}=\lim_{T\to\infty}\frac{1}{T}\int_{0}^{T}V_{\rm bat}(t)^{2}dt$, which for square wave driving amounts to $V_{\rm rms}=\sqrt{V_{0}^{2}+\Delta V^2}$.
Figure~\ref{fig4:cycles}(a) shows a roughly constant enclosed area in {\bf A, B} and {\bf C}, i.e. for $\omega \overline{RC}\le1$. Increasing $\omega \overline{RC}$ beyond unity ({\bf D} and {\bf E}) decreases the enclosed area, and hence decreases the harvested work $W_{\rm har}$. The inset of Fig.~\ref{fig4:cycles}(a) shows that the phase shift $\delta_{\rm max}$ that maximizes $W_{\rm har}$ at a given $\omega$ changes gradually from $\pi/2$ to $\pi$ upon increasing  $\omega \overline{RC}_{1}$. 
This is qualitatively similar to the standard (sinusoidally-driven) RC-circuit which shows an opposite $-\pi/2$ phase shift between $V_{\rm L}$ and $V_{\rm bat}$ upon increasing the angular frequency (see {\it supporting information}).
In the quasi-static $\omega\to0$ limit where $\delta_{\rm max}=\frac{\pi}{2}$, the capacitor goes through four successive steps in which the droplet is compressed at low $V_{\rm cap}$, charged (quickly) at small plate separation separation $L_{0}-\Delta L$, stretched at high $V_{\rm cap}$, followed by a (quick) discharging at maximal separation $L_{0}+\Delta L$. 

There is an interesting analogy to be drawn between this quasi-static cycle and those performed by classic heat engines. The latter have an internal energy $\mathcal{U}(S,V)$ whose differential reads $d\mathcal{U}=\mathsf{T}dS-pdV$, in terms of the entropy $S$ and volume $V$ of a piston at pressure $p$.
A mapping of thermodynamic variables $\mathcal{F}\Leftrightarrow \mathcal{U}$, $-F\Leftrightarrow \mathsf{T}$, $L\Leftrightarrow S$, $V_{\rm cap}\Leftrightarrow -p$ and $Q\Leftrightarrow V$, very much in the spirit of Ref.~\cite{boon2011blue, van2014statistical} where similar reasoning was employed to capacitive mixing processes, shows that the charging cycle is thermodynamically equivalent to an adiabatic compression-isobaric heating-adiabatic expansion-isobaric cooling ``Brayton" cycle \cite{sonntag1998fundamentals}. 

The frequency dependence of the harvested work per cycle is shown in Fig.~\ref{fig4:cycles}(b) for several $\frac{\Delta V}{V_{0}}$. For all oscillation frequencies the introduction of the time-dependent bias potential notably increases the harvested work during a cycle. This effect is most dramatic at low frequencies where square-wave driving results in a finite harvested work, contrary to the constant $V_{\rm bat}(t)=V_{0}$ for which $W_{\rm har}$ vanishes  for decreasing $\omega$. This can also be observed from Fig. \ref{fig4:cycles}(a), where the cycles {\bf A} and {\bf B} enclose a finite area in the charge-potential representation even though $\omega\overline{RC}_{1} \ll 1$. The square-wave driving then leads to an enclosed area of trapezoidal shape, which can be expressed analytically in terms of the charging state $V_{\rm bat}(q)$ at the extremal points of each of the strokes (see {\it supporting information}). 
The resulting expression for the work, eqn~(\ref{seq:approxwork}), increases monotonic with $\Delta L/L_{0}$, and is maximal for $\Delta V/V_{0}=1$. Increasing $\Delta V/V_{0}$ beyond this value leads to a time interval during this cycle where $V(t)$ is negative. The engine delivers negative work in this time interval, because the path, that encloses a finite area in the charge/potential representation reverses its order from anti-clockwise to clockwise. This is supported by Fig.~\ref{fig4:cycles}(b) where the green dotted line at $\Delta V/V_{0}=2$ achieves lower work for all frequencies than the black line at  $\Delta V/V_{0}=1$.
As shown in Fig.~\ref{fig4:cycles}(b), for high oscillation frequencies the work delivered per cycle vanishes. However, the harvested power $P_{\rm har}\equiv W_{\rm har}/T$ (Fig.~\ref{fig4:cycles}(c)) reaches a maximum plateau value in this high-frequency regime. At low frequencies, for which we saw a non-vanishing work output for the electrically driven ($\Delta V>0$) systems, the delivered power per cycle decreases with a smaller slope than the $\Delta V=0$ curve. For typical parameters,  $L_{0}=0.4$mm, and $\Delta L=0.2$mm, the peak in performed work per cycle at $\omega \overline{RC}_{1}\sim 1$ corresponds to a frequency $f\sim50$Hz. The plateau of the delivered power is reached $\omega \overline{RC}_{1}\sim 10$, corresponding to $f\sim500$Hz, a frequency at which hydrodynamic effects will undoubtedly be eminent \cite{moon2013electrical}. Notice however that there are several ways to increase $\omega \overline{RC}_{1}$ without changing $f$. Instead of considering a single droplet, using an array of droplets would increase $\overline{RC}_{1}$. Moreover, using a larger resistance, or using large amplitude oscillations $\frac{\Delta L}{L_{0}}\to 1$ increases $\overline{RC}_{1}$. 

In conclusion, as a first step towards a more realistic treatment of the capacitance of oscillating electrode plates, we determined the droplet profiles of liquid bridges by numerically solving the Young-Laplace equation. 
In this Communication we furthermore showed that the area enclosed in the charge-voltage representation $(q,V_{\rm cap})$ during the steady state of the charging-discharging cycle equals the work harvested by the engine.
This gives a graphic interpretation to work maximization, where the optimal operating regime of the engine is characterized by the largest enclosed areas in this representation.

Varying the average plate separation we found two opposing trends: the engine without external battery power \cite{moon2013electrical, kong2014ionic} thrives at large average plate separation, to be contrasted with the externally-driven droplet engine of  Ref.~\cite{krupenkin2011reverse} which thrives at small average plate separation. We found that both engines benefit from a large oscillation amplitude, in accordance with the experimental findings of Ref.~\cite{moon2013electrical} and \cite{kong2014ionic}.
Explicitly treating surface roughness, to impose pinning of the contact line, is apparently non-essential, since the symmetry between top and bottom electrode can also be broken by gravity and different contact angles rather than different surface heterogeneities. 
However, further work could still include a more systematic treatment of these heterogeneities, to investigate for example the slipping of the contact line. Also, the onset of hydrodynamic effects at high frequency oscillations remains poorly understood.

Our main finding however is that the work and power output of the engine benefit substantially from imposing a time-dependent  bias potential. More specifically, a square-wave battery driving, with an appropriate phase shift with respect to the mechanical driving, leads to quantitative enhancement (by a factor of order 2-5) at high oscillation frequencies, and even to a qualitative enhancement (by several orders of magnitude) at low frequencies, with a non-vanishing work output in this low-frequency regime. We found that a sinusoidal applied potential leads to similar conclusions. Moreover, an optimum was found at $\Delta V/V_{0}=1$, an analytic result that is supported by numerics. We hope that this inspires new experiments to further explore the enhanced conversion of vibrational energy into electric work by a suitably tuned driving.

This work is part of the D-ITP consortium, a program of the 
Netherlands Organisation for Scientific Research (NWO) that is funded by 
the Dutch Ministry of Education, Culture and Science (OCW). 
We acknowledge financial support from an NWO-VICI grant.

\bibliographystyle{rsc}

\bibliography{/Users/mathijsjanssen/Documents/PhD_physics/library/librarymathijs}

\providecommand*{\mcitethebibliography}{\thebibliography}
\csname @ifundefined\endcsname{endmcitethebibliography}
{\let\endmcitethebibliography\endthebibliography}{}
\begin{mcitethebibliography}{22}
\providecommand*{\natexlab}[1]{#1}
\providecommand*{\mciteSetBstSublistMode}[1]{}
\providecommand*{\mciteSetBstMaxWidthForm}[2]{}
\providecommand*{\mciteBstWouldAddEndPuncttrue}
  {\def\EndOfBibitem{\unskip.}}
\providecommand*{\mciteBstWouldAddEndPunctfalse}
  {\let\EndOfBibitem\relax}
\providecommand*{\mciteSetBstMidEndSepPunct}[3]{}
\providecommand*{\mciteSetBstSublistLabelBeginEnd}[3]{}
\providecommand*{\EndOfBibitem}{}
\mciteSetBstSublistMode{f}
\mciteSetBstMaxWidthForm{subitem}
{(\emph{\alph{mcitesubitemcount}})}
\mciteSetBstSublistLabelBeginEnd{\mcitemaxwidthsubitemform\space}
{\relax}{\relax}

\bibitem[Roundy(2005)]{roundy2005effectiveness}
S.~Roundy, \emph{Journal of intelligent material systems and structures}, 2005,
  \textbf{16}, 809--823\relax
\mciteBstWouldAddEndPuncttrue
\mciteSetBstMidEndSepPunct{\mcitedefaultmidpunct}
{\mcitedefaultendpunct}{\mcitedefaultseppunct}\relax
\EndOfBibitem
\bibitem[Beeby \emph{et~al.}(2006)Beeby, Tudor, and White]{beeby2006energy}
S.~P. Beeby, M.~J. Tudor and N.~White, \emph{Measurement science and
  technology}, 2006, \textbf{17}, R175\relax
\mciteBstWouldAddEndPuncttrue
\mciteSetBstMidEndSepPunct{\mcitedefaultmidpunct}
{\mcitedefaultendpunct}{\mcitedefaultseppunct}\relax
\EndOfBibitem
\bibitem[Anton and Sodano(2007)]{anton2007review}
S.~R. Anton and H.~A. Sodano, \emph{Smart materials and Structures}, 2007,
  \textbf{16}, R1\relax
\mciteBstWouldAddEndPuncttrue
\mciteSetBstMidEndSepPunct{\mcitedefaultmidpunct}
{\mcitedefaultendpunct}{\mcitedefaultseppunct}\relax
\EndOfBibitem
\bibitem[Krupenkin and Taylor(2011)]{krupenkin2011reverse}
T.~Krupenkin and J.~A. Taylor, \emph{Nature Communications}, 2011, \textbf{2},
  448\relax
\mciteBstWouldAddEndPuncttrue
\mciteSetBstMidEndSepPunct{\mcitedefaultmidpunct}
{\mcitedefaultendpunct}{\mcitedefaultseppunct}\relax
\EndOfBibitem
\bibitem[Meninger \emph{et~al.}(2001)Meninger, Mur-Miranda, Amirtharajah,
  Chandrakasan, and Lang]{meninger2001vibration}
S.~Meninger, J.~O. Mur-Miranda, R.~Amirtharajah, A.~P. Chandrakasan and J.~H.
  Lang, \emph{Very Large Scale Integration (VLSI) Systems, IEEE Transactions
  on}, 2001, \textbf{9}, 64--76\relax
\mciteBstWouldAddEndPuncttrue
\mciteSetBstMidEndSepPunct{\mcitedefaultmidpunct}
{\mcitedefaultendpunct}{\mcitedefaultseppunct}\relax
\EndOfBibitem
\bibitem[Boisseau \emph{et~al.}(2012)Boisseau, Despesse, and
  Seddik]{boisseau2012electrostatic}
S.~Boisseau, G.~Despesse and B.~A. Seddik, in \emph{Small-Scale Energy
  Harvesting}, ed. D.~M. Lallart, InTech, 2012, ch. Electrostatic conversion
  for vibration energy harvesting\relax
\mciteBstWouldAddEndPuncttrue
\mciteSetBstMidEndSepPunct{\mcitedefaultmidpunct}
{\mcitedefaultendpunct}{\mcitedefaultseppunct}\relax
\EndOfBibitem
\bibitem[Brogioli(2009)]{brogioli2009extracting}
D.~Brogioli, \emph{Physical Review Letters}, 2009, \textbf{103}, 058501\relax
\mciteBstWouldAddEndPuncttrue
\mciteSetBstMidEndSepPunct{\mcitedefaultmidpunct}
{\mcitedefaultendpunct}{\mcitedefaultseppunct}\relax
\EndOfBibitem
\bibitem[Hamelers \emph{et~al.}(2013)Hamelers, Schaetzle, Paz-Garc{\'\i}a,
  Biesheuvel, and Buisman]{hamelers2013harvesting}
H.~Hamelers, O.~Schaetzle, J.~Paz-Garc{\'\i}a, P.~Biesheuvel and C.~Buisman,
  \emph{Environmental Science \& Technology Letters}, 2013, \textbf{1},
  31--35\relax
\mciteBstWouldAddEndPuncttrue
\mciteSetBstMidEndSepPunct{\mcitedefaultmidpunct}
{\mcitedefaultendpunct}{\mcitedefaultseppunct}\relax
\EndOfBibitem
\bibitem[H{\"a}rtel \emph{et~al.}(2015)H{\"a}rtel, Janssen, Weingarth, Presser,
  and van Roij]{hartel2015heat}
A.~H{\"a}rtel, M.~Janssen, D.~Weingarth, V.~Presser and R.~van Roij,
  \emph{Energy \& Environmental Science}, 2015, \textbf{8}, 2396--2401\relax
\mciteBstWouldAddEndPuncttrue
\mciteSetBstMidEndSepPunct{\mcitedefaultmidpunct}
{\mcitedefaultendpunct}{\mcitedefaultseppunct}\relax
\EndOfBibitem
\bibitem[Janssen \emph{et~al.}(2014)Janssen, H{\"a}rtel, and van
  Roij]{Janssen:2014aa}
M.~Janssen, A.~H{\"a}rtel and R.~van Roij, \emph{Physical Review Letters},
  2014, \textbf{113}, 268501\relax
\mciteBstWouldAddEndPuncttrue
\mciteSetBstMidEndSepPunct{\mcitedefaultmidpunct}
{\mcitedefaultendpunct}{\mcitedefaultseppunct}\relax
\EndOfBibitem
\bibitem[Ahualli \emph{et~al.}(2014)Ahualli, Fern{\'a}ndez, Iglesias, Delgado,
  and Jim{\'e}nez]{ahualli2014temperature}
S.~Ahualli, M.~M. Fern{\'a}ndez, G.~Iglesias, {\'A}.~V. Delgado and M.~L.
  Jim{\'e}nez, \emph{Environmental Science \& Technology}, 2014, \textbf{48},
  12378--12385\relax
\mciteBstWouldAddEndPuncttrue
\mciteSetBstMidEndSepPunct{\mcitedefaultmidpunct}
{\mcitedefaultendpunct}{\mcitedefaultseppunct}\relax
\EndOfBibitem
\bibitem[Miyazaki \emph{et~al.}(2003)Miyazaki, Tanaka, Ono, Nagano, Ohkubo,
  Kawahara, and Yano]{miyazaki2003electric}
M.~Miyazaki, H.~Tanaka, G.~Ono, T.~Nagano, N.~Ohkubo, T.~Kawahara and K.~Yano,
  Low Power Electronics and Design, 2003. ISLPED'03. Proceedings of the 2003
  International Symposium on, 2003, pp. 193--198\relax
\mciteBstWouldAddEndPuncttrue
\mciteSetBstMidEndSepPunct{\mcitedefaultmidpunct}
{\mcitedefaultendpunct}{\mcitedefaultseppunct}\relax
\EndOfBibitem
\bibitem[Moon \emph{et~al.}(2013)Moon, Jeong, Lee, and Pak]{moon2013electrical}
J.~K. Moon, J.~Jeong, D.~Lee and H.~K. Pak, \emph{Nature Communications}, 2013,
  \textbf{4}, 1487\relax
\mciteBstWouldAddEndPuncttrue
\mciteSetBstMidEndSepPunct{\mcitedefaultmidpunct}
{\mcitedefaultendpunct}{\mcitedefaultseppunct}\relax
\EndOfBibitem
\bibitem[Kong \emph{et~al.}(2014)Kong, Cao, He, Yu, Ma, He, Lu, Zhang, and
  Deng]{kong2014ionic}
W.~Kong, P.~Cao, X.~He, L.~Yu, X.~Ma, Y.~He, L.~Lu, X.~Zhang and Y.~Deng,
  \emph{RSC Advances}, 2014, \textbf{4}, 19356--19361\relax
\mciteBstWouldAddEndPuncttrue
\mciteSetBstMidEndSepPunct{\mcitedefaultmidpunct}
{\mcitedefaultendpunct}{\mcitedefaultseppunct}\relax
\EndOfBibitem
\bibitem[Batchelor(2000)]{batchelor2000introduction}
G.~K. Batchelor, \emph{An introduction to fluid dynamics}, Cambridge university
  press, 2000\relax
\mciteBstWouldAddEndPuncttrue
\mciteSetBstMidEndSepPunct{\mcitedefaultmidpunct}
{\mcitedefaultendpunct}{\mcitedefaultseppunct}\relax
\EndOfBibitem
\bibitem[Kornyshev(2007)]{kornyshev2007double}
A.~A. Kornyshev, \emph{The Journal of Physical Chemistry B}, 2007,
  \textbf{111}, 5545--5557\relax
\mciteBstWouldAddEndPuncttrue
\mciteSetBstMidEndSepPunct{\mcitedefaultmidpunct}
{\mcitedefaultendpunct}{\mcitedefaultseppunct}\relax
\EndOfBibitem
\bibitem[H{\"a}rtel \emph{et~al.}(2015)H{\"a}rtel, Janssen, Samin, and van
  Roij]{hartel2015fundamental}
A.~H{\"a}rtel, M.~Janssen, S.~Samin and R.~van Roij, \emph{Journal of Physics:
  Condensed Matter}, 2015, \textbf{27}, 194129\relax
\mciteBstWouldAddEndPuncttrue
\mciteSetBstMidEndSepPunct{\mcitedefaultmidpunct}
{\mcitedefaultendpunct}{\mcitedefaultseppunct}\relax
\EndOfBibitem
\bibitem[Klarman \emph{et~al.}(2011)Klarman, Andelman, and
  Urbakh]{klarman2011model}
D.~Klarman, D.~Andelman and M.~Urbakh, \emph{Langmuir}, 2011, \textbf{27},
  6031--6041\relax
\mciteBstWouldAddEndPuncttrue
\mciteSetBstMidEndSepPunct{\mcitedefaultmidpunct}
{\mcitedefaultendpunct}{\mcitedefaultseppunct}\relax
\EndOfBibitem
\bibitem[Buehrle \emph{et~al.}(2003)Buehrle, Herminghaus, and
  Mugele]{buehrle2003interface}
J.~Buehrle, S.~Herminghaus and F.~Mugele, \emph{Physical Review Letters}, 2003,
  \textbf{91}, 086101\relax
\mciteBstWouldAddEndPuncttrue
\mciteSetBstMidEndSepPunct{\mcitedefaultmidpunct}
{\mcitedefaultendpunct}{\mcitedefaultseppunct}\relax
\EndOfBibitem
\bibitem[Boon and van Roij(2011)]{boon2011blue}
N.~Boon and R.~van Roij, \emph{Molecular Physics}, 2011, \textbf{109},
  1229--1241\relax
\mciteBstWouldAddEndPuncttrue
\mciteSetBstMidEndSepPunct{\mcitedefaultmidpunct}
{\mcitedefaultendpunct}{\mcitedefaultseppunct}\relax
\EndOfBibitem
\bibitem[van Roij(2014)]{van2014statistical}
R.~van Roij, in \emph{Electrostatics of Soft and Disordered Matter}, ed. A.~N.
  . R.~P. David S.~Dean, Jure~Dobnikar, Singapore: Pan Stanford Publishing,
  2014, ch. Statistical thermodynamics of supercapacitors and blue engines, p.
  263\relax
\mciteBstWouldAddEndPuncttrue
\mciteSetBstMidEndSepPunct{\mcitedefaultmidpunct}
{\mcitedefaultendpunct}{\mcitedefaultseppunct}\relax
\EndOfBibitem
\bibitem[Sonntag \emph{et~al.}(1998)Sonntag, Borgnakke, Van~Wylen, and
  Van~Wyk]{sonntag1998fundamentals}
R.~E. Sonntag, C.~Borgnakke, G.~J. Van~Wylen and S.~Van~Wyk, \emph{Fundamentals
  of thermodynamics}, Wiley New York, 1998, vol.~6\relax
\mciteBstWouldAddEndPuncttrue
\mciteSetBstMidEndSepPunct{\mcitedefaultmidpunct}
{\mcitedefaultendpunct}{\mcitedefaultseppunct}\relax
\EndOfBibitem
\end{mcitethebibliography}

\newpage
\section{Supporting Information}

\setcounter{equation}{0}
\setcounter{figure}{0}
\setcounter{table}{0}
\makeatletter 
\renewcommand{\thefigure}{S\@arabic\c@figure}
\renewcommand{\thetable}{S\@arabic\c@table}
\renewcommand{\theequation}{S\@arabic\c@equation}
\makeatother
\subsection{Traditional RC-circuit}

\begin{figure}[h!]
\centering
\includegraphics[width=8.0cm]{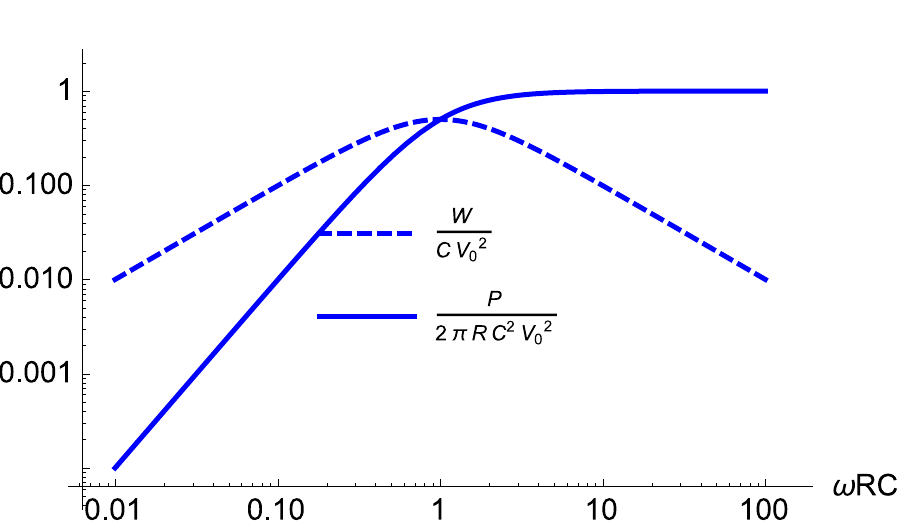}
\caption{\scriptsize The work and power performed over a load resistor in a traditional RC-circuit.}\label{fig:traditionalRC}
\end{figure}

For comparison, we quickly recap some elementary results for traditional RC-circuits.
The formal solution to eqn~(\ref{eq:kirchoffgeneral}) reads
\begin{eqnarray}\label{formalsolution}
q(t)&=&q_{0}e^{-\int_{0}^{t}\frac{1}{RC(t')}dt'} \nonumber\\
&&+e^{-\int_{0}^{t}\frac{1}{RC(t')}dt'} \int^{t}_{0} dt' \frac{V(t')}{R} e^{\int_{0}^{t'}\frac{1}{RC(t'')}dt''},
\end{eqnarray}
which can for instance be found with variation of constants.
A traditional RC-circuit has a constant capacitance $C(t)=C$, such that eqn~(\ref{eq:kirchoffgeneral}) becomes a first order linear time-invariant theory,
for which eqn~(\ref{formalsolution}) simplifies to the well-known expression
\begin{eqnarray}
q(t)&=&q_{0}e^{-\frac{t}{RC}} + \int^{t}_{0} dt' \frac{V(t')}{R} e^{\frac{t'-t}{RC}}.
\end{eqnarray}
In the case of a sinusoidal driving $V(t)=V_{0}\sin{\omega t}$, 
we then find the potential over the load 
\begin{eqnarray}
V_{\rm L}&=& \frac{RCi\omega}{1+RCi\omega}V_{0},
\end{eqnarray}
The phase angle between $V_{\rm L}$ and $V_{0}$ amounts to $\phi_{\rm L}=\tan^{-1}\left(\frac{1}{\omega RC}\right)$, from which we see that that there is a $\pi/2$ phase shift when increasing $\omega$ from $\omega\to0$ ($\phi_{\rm L}=\pi/2$) to  $\omega\to\infty$ ($\phi_{\rm L}=0$). 
In Fig.~\ref{fig:traditionalRC} we show the power $P=\frac{|V_{\rm L}|^{2}}{R}$ dissipated over the resistor and
relatedly, the energy $W=P\cdot T$ dissipated over the load during one cycle. The power drops to its half-value at the {\it cut-off} angular frequency $\omega RC=1$. Relatedly, the work is seen to have a maximum $W(\omega =1/RC)=\frac{1}{2}CV_{0}^{2}$ located at the same frequency.

\subsubsection{Variable capacitance RC-circuit}
The analytic solution Krupenkin and Taylor \cite{krupenkin2011reverse} present for the case of constant potential $V(t)=V_{0}$ with a sinusoidally varying capacitance $C_{t/b}(t)=C_{0}(1+\cos \omega t$), can be found by plugging in into eqn~(\ref{formalsolution}). However, for the liquid bridge engine we are interested in, this expression for the capacitance is not correct since a vanishing capacitance would imply a broken bridge. 

\subsection{Dimensionless equations }
The Kirchoff equation eqn~(\ref{eq:kirchoffgeneral}) is brought in dimensionless form by writing $q(t)=\hat{q}(\hat{t}) V_{0}C_{0}$, $t=\hat{t}\omega^{-1}$, $C_{b/t}(t)=C_{b/t}(0)\hat{C}_{b/t}(t)$ and $C_{tot}=C_{0}\hat{C}_{tot}$, and $V(t)=V_{0}\hat{V}(t)$. The choice of $C_{0}\equiv C_{tot}(0)$ fixes $\hat{C}_{tot}^{-1}=\hat{C}_{b}^{-1}\left(1+\frac{C_{b}(0)}{C_{t}(0)}\right)^{-1}+\hat{C}_{t}^{-1}\left(1+\frac{C_{t}(0)}{C_{b}(0)}\right)^{-1}$.  
Defining $\alpha=\omega RC_{0}$, the Kirchoff equation reads
\begin{eqnarray}\label{eq:kirchoffdimensionlessII}
\alpha\frac{d\hat{q}(t)}{d\hat{t}}+\frac{\hat{q}(\hat{t})}{\hat{C}_{tot}(\hat{t})}&=&\hat{V}(t)\\
\hat{q}(0)&=&1
\end{eqnarray}
The energy harvested over the load, written in terms of the dimensionless variables, reads
\begin{eqnarray}
W_{\rm L}&=&\int^{\hat{t}(T)}_{0} d\hat{t} \frac{d\hat{q}(\hat{t})}{d\hat{t}}(\hat{V}(t)-\frac{\hat{q}(\hat{t})}{\hat{C}_{tot}})V_{0}^{2}C_{0}
\end{eqnarray}

In a set-up {\bf without external power source} the natural voltage scale is set by $\frac{Q_{b}}{C_{b}(0)}$, this equation is brought in dimensionless form by writing $q(t)=\hat{q}(\hat{t}) Q_{b}$, $t=\hat{t}\omega^{-1}$, $C_{tot}=C_{0}\hat{C}_{tot}$, and $C_{b/t}(t)=C_{b/t}(0)\hat{C}_{b/t}(t)$ which gives
\begin{eqnarray}\label{eq:Kirchoffasymdimless2}
\alpha\frac{d\hat{q}(\hat{t})}{d\hat{t}}+\frac{\hat{q}(\hat{t}) }{\hat{C}_{tot}(\hat{t})}&=&\frac{C_{0}}{C_{b}(0)}\left(\frac{1}{\hat{C}_{b}(\hat{t})}-\frac{1}{\hat{C}_{t}(\hat{t})}\right).
\end{eqnarray}

\subsection{Static limit of square wave driving}
\begin{figure}
\centering
\includegraphics[width=6.0cm]{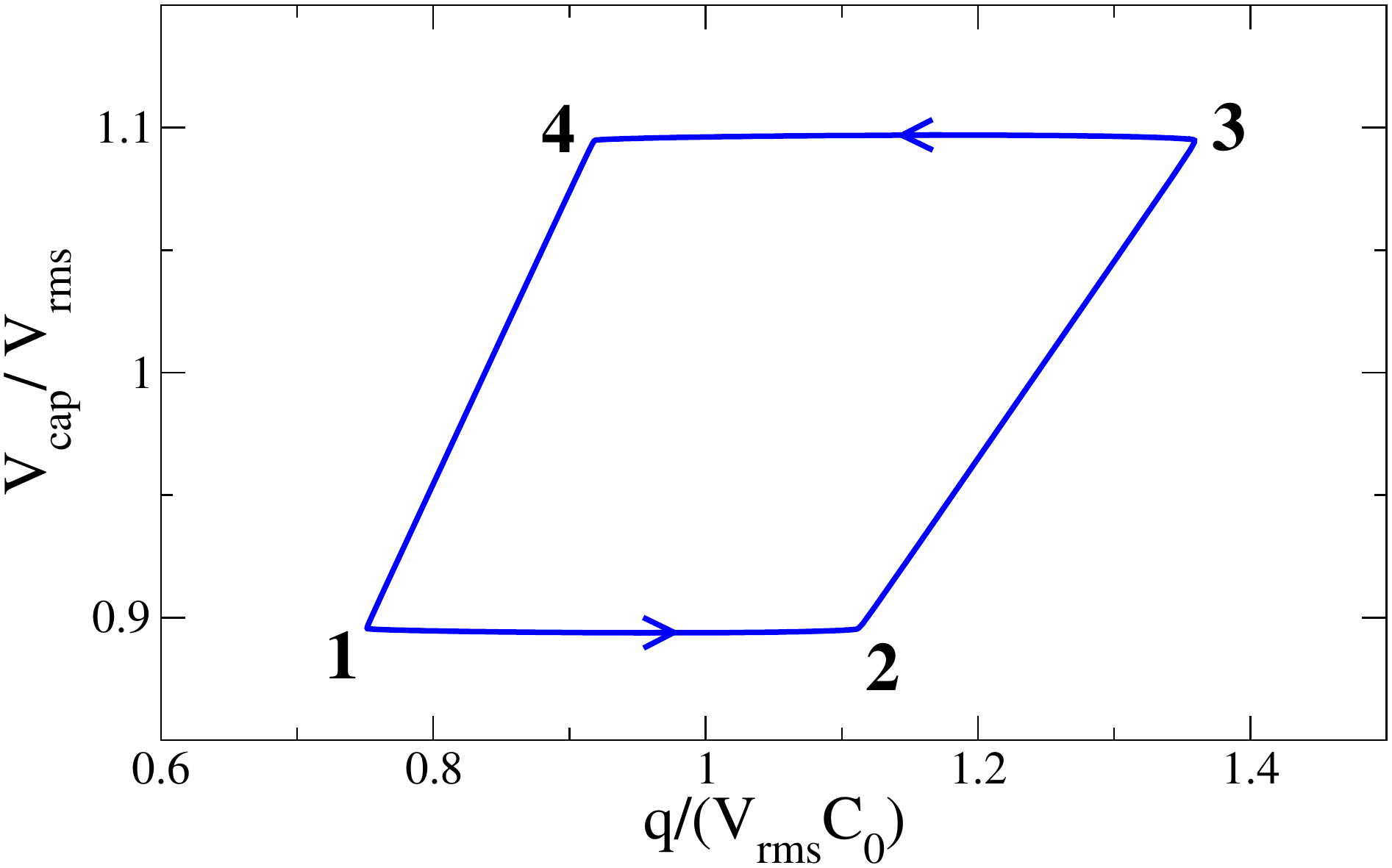}
\caption{\scriptsize{Definition of the state-points}}
\end{figure}
The square wave driving gives a trapezoidal enclosed area such that the harvested work amounts to
\begin{equation}
W_{\omega\to0}=\left(\left(Q_{2}-Q_{1}\right)+\left(Q_{3}-Q_{4}\right)\right)\Delta V
\end{equation}
The charges can be expressed as $Q_{1}=V_{l}C_{l}$,  $Q_{2}=V_{l}C_{h}$,  $Q_{3}=V_{h}C_{h}$, and  $Q_{4}=V_{h}C_{h}$ with $l$ low and $h$ high. Combining gives
\begin{equation}
W_{\omega\to0}=\left(V_{l}\left(C_{h}-C_{l}\right)+V_{h}\left(C_{h}-C_{l}\right)\right)\Delta V
\end{equation}
Writing $\Delta C\equiv C_{h}-C_{l}$ and because $V_{l}=V_{0}-\Delta V$ and $V_{h}=V_{0}+\Delta V$ we find
\begin{equation}
W_{\omega\to0}=2\Delta CV_{0}\Delta V
\end{equation}
We normalize this to $C_{0}V_{\rm rms}^2$, using that the root mean square potential 
\begin{equation}
\frac{W_{\omega\to0}}{C_{0}V_{\rm rms}^2}=2\frac{\Delta C}{C_{0}}\frac{\frac{\Delta V}{V_{0}}}{1+\left(\frac{\Delta V}{V_{0}}\right)^{2}},
\end{equation}
which is maximal at $\frac{\Delta V}{V_{0}}=1$. For $L\ll l_{c}$ we can use that $A\sim\frac{1}{L}$ together with eqn~(\ref{eq:capacitances}) to find
\begin{equation}
\frac{\Delta C}{C_{0}}=2\frac{\frac{\Delta{L}}{L_{0}}}{1-\left(\frac{\Delta{L}}{L_{0}}\right)^{2}},
\end{equation}
which gives the final result
\begin{equation}\label{seq:approxwork}
\frac{W_{\omega\to0}}{C_{0}V_{\rm rms}^2}=4\frac{\frac{\Delta{L}}{L_{0}}}{1-\left(\frac{\Delta{L}}{L_{0}}\right)^{2}}\frac{\frac{\Delta V}{V_{0}}}{1+\left(\frac{\Delta V}{V_{0}}\right)^{2}}.
\end{equation}

\end{document}